\documentclass{article} 
\usepackage{ijcai23}

\usepackage{times}
\usepackage{soul}
\usepackage{url}
\usepackage[hidelinks]{hyperref}
\usepackage[utf8]{inputenc}
\usepackage[small]{caption}
\usepackage{graphicx}
\usepackage{amsmath}
\usepackage{amsthm}
\usepackage{amsfonts}
\usepackage{mathtools}
\usepackage{booktabs}
\usepackage{algorithm}
\usepackage{algorithmic}
\usepackage[switch]{lineno}
\usepackage{amssymb,enumitem,xfrac}

\usepackage{enumitem}
\usepackage{color}

\newcommand{\expected}{\mathbb{E}}

\newcommand{\R}{\mathcal{R}}

\newtheorem{theorem}{Theorem}[section]
\newtheorem{lemma}[theorem]{Lemma}

\newtheorem{definition}[theorem]{Definition}

\newtheorem{remark}{Remark}

\graphicspath{{figures/}}
\DeclareGraphicsExtensions{.pdf,.jpeg,.png,.jpg}

\newcommand{\prob}{\textsc{MobileVaccClinic}}

\title{Differentially Private Partial Set Cover with Applications to Facility Location}
\author{
George Z. Li$^1$, Dung Nguyen$^2$,\And Anil Vullikanti$^2$
\affiliations
$^1$University of Maryland\\
$^2$Biocomplexity Institute and Initiative, and Department of Computer Science, University of Virginia\\
\emails
gzli929@gmail.com,
dungn@virginia.edu,
vsakumar@virginia.edu
}

\begin{document}

\maketitle

\begin{abstract}
Set Cover is a fundamental problem in combinatorial optimization which has been studied for many decades due to its various applications across multiple domains.
In many of these domains, the input data consists of locations, relationships, and other sensitive information of individuals which may leaked due to the set cover output.
Attempts have been made to design privacy-preserving algorithms to solve the Set Cover under privacy constraints.
Under differential privacy, it has been proved that the Set Cover problem has strong impossibility results and no explicit forms of the output can be released to the public.

In this work, we observe that these hardness results dissolve when we turn to the Partial Set Cover problem, where we only need to cover a $\rho\in(0,1)$ fraction of the elements.
We show that this relaxation enables us to avoid the impossibility results, and give the first algorithm which outputs an explicit form of set cover with non-trivial utility guarantees under differential privacy.
Using our algorithm as a subroutine, we design a differentially private bicriteria algorithm to solve a recently proposed facility location problem for vaccine distribution which generalizes the $k$-supplier with outliers.
Our analysis shows that relaxing the covering requirement to serve only a $\rho\in(0,1)$ fraction of the population/universe also allows us to circumvent the inherent hardness of $k$-supplier and give the first non-trivial guarantees.
\end{abstract}

\section{Introduction}

Data privacy is a fundamental challenge in many real world applications of data-driven decision making where there is a risk of inadvertently revealing private information. Differential privacy, introduced in~\cite{DBLP:conf/tcc/DworkMNS06}, has emerged as a widely accepted formalization of privacy, which gives rigorous parameterized guarantees on the privacy loss while simultaneously enabling non-trivial utility in algorithmic and statistical analysis. Differential privacy is defined in terms of datasets which differ by one individual, called neighboring datasets, and requires that the output of a mechanism is (approximately) indistinguishable when run on any two neighboring datasets. Formally, it is defined as:
\begin{definition}
$M:\mathcal{X}^n\to\mathcal{Y}$ is $(\epsilon,\delta)$-differentially private if for any neighboring datasets $X,X^\prime \in\mathcal{X}^n$ and $S\subseteq\mathcal{Y}$,
\begin{align}
    \Pr[M(X)\in S]\le \exp(\epsilon)\Pr[M(X^\prime)\in S] + \delta.\nonumber
\end{align}
If $\delta=0$, we say $M$ is $\epsilon$-differentially private.
\end{definition}
Differentially private algorithms have now been developed for a number of problems ranging from statistics~\cite{DBLP:conf/nips/Canonne0MUZ20,DBLP:journals/corr/abs-2106-13329}, machine learning and deep learning~\cite{10.1145/3219819.3220076,DBLP:journals/corr/abs-2102-06062}, social network analysis~\cite{DBLP:conf/stoc/NissimRS07,DBLP:conf/icdm/HayLMJ09,DBLP:journals/pvldb/KarwaRSY11}, and combinatorial optimization~\cite{DBLP:conf/icml/MitrovicB0K17,DBLP:conf/nips/EsencayiGLW19,DBLP:conf/icml/NguyenV21}. See~\cite{dwork_and_roth,DBLP:books/sp/17/Vadhan17} for a survey on the techniques used.

In this work, we consider a fundamental problem in combinatorial optimization: the Set Cover problem~\cite{williamson_shmoys_2011}, which involves choosing the smallest subset of a set system $\mathcal{S}=\{S_1,\ldots,S_m\}\subset 2^{\mathcal{U}}$ that covers a universe $\mathcal{U}=\{u_1,\ldots,u_n\}$.
In many settings, the elements of the universe $\mathcal{U}$ are private (e.g., clients wish to be private in facility location problems).
\cite{gupta2009differentially} first studied the problem of Set Cover with privacy, and showed that outputting an explicit solution to Set Cover has strong hardness results, even for the special case of Vertex Cover: any differentially private algorithm must output a set cover of size $m-1$ with probability 1 on any input, a useless result. 
As a result, the authors designed a mechanism which outputs an implicit set cover via a privacy-preserving set of instruction for the elements to reconstruct the set cover. While the implicit solutions are useful in some limited settings, it cannot replace the \emph{explicit} solutions needed in many important applications, such as public health~\cite{DBLP:conf/soda/EubankKMSW04,li2022deploying}. In particular, explicit solutions are necessary when using a Set Cover algorithm as a subroutine when solving a more complicated problem. As a result, we turn to the Partial Set Cover problem, where we only need to cover a $\rho$-fraction of the elements in $\mathcal{U}$, for $\rho\in(0,1)$. 
Due to the space limit, we present our core results in the main paper and defer some analyses and experimental results to the Appendix. We maintain a full, updated version of this work here\footnote{\url{https://arxiv.org/abs/2207.10240}}.

Our primary contributions are:
\begin{itemize}
    \item We observe that the impossibility results for outputting an explicit set cover under differential privacy are alleviated when considering the Partial Set Cover problem. When the number sets isn't too large (i.e., $m=O(n)$), we give a $O(\log^2(m)\log(1/\delta)/\epsilon(1-\rho))$-approximation algorithm (see Theorem \ref{thm:greedy_main}). Alternatively, when the optimal partial set cover isn't too large (i.e., $\text{OPT}\lesssim \frac{n\epsilon}{\log^3{n}\log(1/\delta)}$), we give a $O(\log(\frac{1}{(1-\rho)}))$-pseudo approximation algorithm (see Theorem \ref{thm:cov}). Note that both of our guarantees break down as $\rho\to 1$.
    \item To illustrate the importance of explicit solutions, we use our differentially private Partial Set Cover algorithm as a subroutine to give a differentially private approximation algorithm for a vaccine distribution problem which in particular, generalizes $k$-supplier with outliers. We emphasize that this is the first differentially private algorithm for $k$-supplier type problems with non-trivial approximation guarantees, which was thought to be impossible due to the high sensitivity of min-max objectives.
    \item
    Finally, we evaluate the our private algorithms on real life datasets for set cover and vaccine distribution, and observe a good privacy-utility tradeoff: the empirical approximation factors are much better than the worst case bounds we give in our theorem statements. We view the vaccine distribution problem as a concrete practical contribution of our work (see Figure 1 for an example).
\end{itemize}

\begin{figure}[h!]
    \centering
    \includegraphics[width=\columnwidth,keepaspectratio]{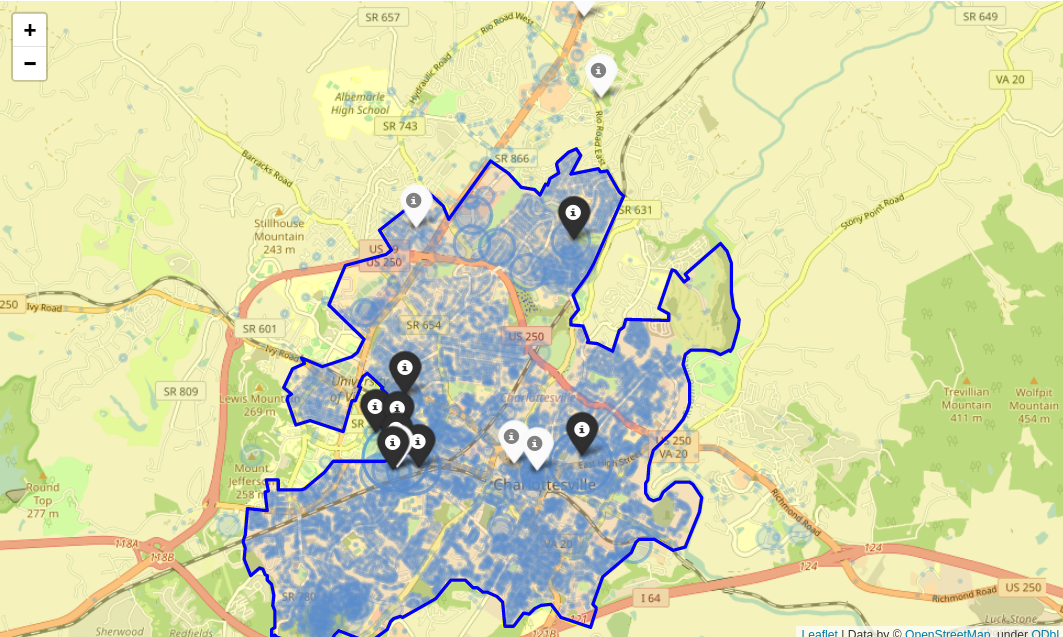}
    \caption{Visualization of locations of chosen facilities. The black and white spots indicate locations chosen by the non-private and private algorithms, respectively. As we see, the private algorithm yields qualitatively different distributions when compared to the non-private algorithm while not losing too much in objective value (see Section \ref{sec:exp}).}
    \label{fig:Cville}
  \end{figure}

\subsection{Related Work}
The Set Cover problem and its various generalizations have been studied by combinatorial optimization community for several decades~\cite{DBLP:journals/combinatorica/Wolsey82,DBLP:conf/stoc/AlonAA03}. For the simplest version, there exists a greedy algorithm which achieves a $(\log{n}+1)$-approximation which is best possible unless P$=$NP~\cite{v011a007}.
\cite{gupta2009differentially} first considered the Set Cover problem under differential privacy, showing impossibility results of outputting explicit set covers. They then gave approximation algorithms via outputting implicit set covers, which we argue is insufficient for many applications. The only other work which outputs explicit set covers under differential privacy is~\cite{DBLP:conf/icalp/HsuRRU14}, which approach the set cover problem via private linear programming. They give approximation guarantees but ignore $O(\text{OPT}^2\cdot\text{polylog}(n,1/\delta))$ elements. One of our algorithms also has guarantees of this form, but ignores only $\rho n$ elements, which is better when the size of the optimal set cover is larger. We also give a true approximation for Partial Set Cover, which is the first in the literature.

Also directly related to our work are the differentially private facility location problems, which
\cite{gupta2009differentially} also first considered. For the uniform facility location problem, they showed that a $\Omega(\sqrt{n})$-approximation is needed under differential privacy, an essentially useless result, and devised a way to implicitly output the facilities.~\cite{DBLP:conf/nips/EsencayiGLW19} built upon their work, improving the approximation guarantees to $O(\frac{\log{n}}{\epsilon})$ for general metrics.~\cite{DBLP:conf/aistats/Cohen-AddadEFG022} considered the facility location problem under the local differential privacy model, gave tight approximation algorithms up to polynomial factors of $\epsilon$. A particular interesting quality of their algorithms is that it extends to non-uniform facility location.~\cite{gupta2009differentially} also considered the $k$-median problem, and developed approximation algorithms which guaranteed that the service cost is at most $6\cdot \text{OPT}+O(\frac{k^2\log^2{n}}{\epsilon})$. Since then, there has been an abundance of work on this problem improving the approximation guarantees, practical performance, and efficiency of the differentially private algorithms~\cite{DBLP:conf/icml/BalcanDLMZ17,NEURIPS2020_299dc35e,DBLP:conf/isit/BlockiGM21,DBLP:conf/aaai/JonesNN21,DBLP:journals/corr/abs-2206-08646}. Despite the abundance of work on facility location-type problems, $k$-supplier remains untouched; our work is the first to overcome impossibility results and give approximation guarantees for the problem.

\subsection{Differential Privacy Background}
In our algorithms, we will make extensive use of the following basic mechanisms and properties, the proofs of which can be found in \cite{dwork_and_roth}. The post-processing and composition properties enable us to easily combine smaller differentially private mechanisms to create a more complicated one:

\begin{theorem}
Let $\mathcal{M}_1:\mathcal{X}^n\to\mathcal{Y}_1$ and $\mathcal{M}_2:\mathcal{X}^n\to\mathcal{Y}_2$ be $(\epsilon_1,\delta_1)$ and $(\epsilon_2,\delta_2)$-differentially private algorithms, respectively. The following properties hold:
\begin{itemize}[noitemsep]
    \item \textbf{post-processing}: Let $f:\mathcal{Y}_1\to\mathcal{Z}$ be an arbitrary (potentially randomized) mapping. Then $f\circ \mathcal{M}_1:\mathcal{X}^n\to\mathcal{Z}$ is $(\epsilon_1,\delta_1)$-differentially private.
    \item \textbf{composition}: Let $\mathcal{M}:\mathcal{X}^n\to\mathcal{Y}_1\times\mathcal{Y}_2$ be defined as $\mathcal{M}=(\mathcal{M}_1,\mathcal{M}_2)$. Then $\mathcal{M}$ is $(\epsilon_1+\epsilon_2,\delta_1+\delta_2)$-differentially private.
\end{itemize}
\end{theorem}

We next state the Laplace Mechanism which, by adding Laplace noise, provides a simple way to privately output a statistic that depends on a private database:

\begin{theorem}
Given a function $f:\mathcal{X}^n\to\mathbb{R}^k$, the $\ell_1$-sensitivity is defined as $\Delta_f=\max_{x\sim x^\prime}\|f(x)-f(x^\prime)\|_1$. The Laplace Mechanism, given the function $f:\mathcal{X}^n\to\mathbb{R}^k$, outputs $f(x)+(Y_1,\ldots,Y_k)$, where $Y_i$ are i.i.d. random variables drawn from $\text{Lap}(\Delta_f/\epsilon)$. We claim the Laplace Mechanism is $\epsilon$-differentially private.
\end{theorem}

Finally, we state the Exponential Mechanism which approximately optimizes the utility function over some set of candidates choices $\mathcal{R}$ while preserving privacy:

\begin{definition}
Given a utility function $u:\mathcal{X}^n\times\mathcal{R}\to\mathbb{R}$, let $\Delta_u=\max_{r\in\mathcal{R}}\max_{x\sim x^\prime}|u(x,r)-u(x^\prime,r)|$ be the sensitivity of $u$, where $x,x^\prime$ are neighboring datasets. The exponential mechanism $M(x,u,\mathcal{R})$ outputs an element $r\in\mathcal{R}$ with probability $\propto\exp(\frac{\epsilon u(x,r)}{2\Delta_u})$.
\end{definition}

\begin{theorem}
The exponential mechanism is $\epsilon$-differentially private. Furthermore, if we fix a dataset $x$ and let $\text{OPT}=\max_{r\in\mathcal{R}}u(x,r)$, we have $\Pr[u(x,M(x,u,\mathcal{R}))\le \text{OPT}-\frac{2\Delta_u}{\epsilon}(\ln|\mathcal{R}|+t)]\le e^{-t}$.
\end{theorem}

\section{Differentially Private Partial Set Cover} \label{sec:cover}
Formally, we wish to solve the following problem:
\begin{definition}
Let $\mathcal{U}=\{u_1,\ldots,u_n\}$ be the universe of elements and let $\mathcal{S}=\{S_1,\ldots,S_m\}$ be a set system where each $S_i\subseteq \mathcal{U}$ and $\bigcup_{i=1}^{m}S_i=\mathcal{U}$.  Finally, let $\rho<1$ be the covering requirement. The Partial Set Cover problem asks for the minimal size subset $\{\pi_1,\ldots,\pi_k\}$ of $\mathcal{S}$ such that $|\bigcup_{i=1}^{k}\pi_i|\ge \rho |\mathcal{U}|$ (i.e., the subset covers a $\rho$ fraction of $\mathcal{U}$).
\end{definition}

For differential privacy, we consider two Partial Set Cover instances neighboring if the set systems $(\mathcal{U}_1,\mathcal{S}_1)$ and $(\mathcal{U}_2,\mathcal{S}_2)$ are two neighboring datasets if they differ by exactly one element $u$ in the universe and the sets $S_{i1}\in\mathcal{S}_1$ and $S_{i2}\in\mathcal{S}_2$ differ only by $u$ (or are the same). Note that our definition of privacy can recover the one considered in \cite{gupta2009differentially} and \cite{DBLP:conf/icalp/HsuRRU14} as a special case, but is in general stronger. We elaborate on this subtlety further in the Appendix. 

At first glance, our (and \cite{gupta2009differentially}'s) definition of neighboring datasets for (Partial) Set Cover may not be entirely intuitive.
As an example of an application where such a privacy definition makes sense, let's consider the problem studied by \cite{DBLP:conf/soda/EubankKMSW04} of placing sensors in people-location graphs to detect the spread of a disease. Formally, we have a bipartite graph where nodes in one part of the graph represent locations and nodes in the other part represent people. An edge exists between a person and a location if the person visits that location. We wish to place sensors at the locations so that infected individuals can be detected. Since placing sensors is an expensive process, we wish to place the fewest sensors as possible to cover a $\rho$-fraction of the population. This can be formulated as a Partial Set Cover problem. For this example, our notion of privacy corresponds to node privacy for people in the graph. In addition to this example, we show in the following section that our privacy definition here also lines up nicely with client-privacy in facility location problems when using Partial Set Cover to solve $k$-supplier with outliers.

In the following subsections, we'll present our algorithms for differentially private Partial Set Cover. Since Partial Set Cover is NP-Hard, we will focus on giving approximation algorithms while preserving differential privacy. We say an algorithm gives an $\alpha$-approximation to the optimal solution if it outputs a partial set cover of size at most $\alpha\cdot\text{OPT}$, where $\text{OPT}$ is the size of an optimal partial set cover.

\subsection{A Private Variant of the Greedy Algorithm}

The general outline of our first algorithm is as follows: we use a private version of the classical greedy algorithm for Partial Set Cover to output a permutation of the sets $\pi_1,\ldots,\pi_m$. Then, we use an offline implementation of the AboveThreshold mechanism to choose a threshold $k$ such that $\pi_1,\ldots,\pi_k$ covers a $\rho$ fraction of the elements.

\begin{algorithm}
\caption{$\mathcal{M}_{PartialSetCover}(\mathcal{U},\mathcal{S},\rho,\epsilon,\delta)$}
\begin{algorithmic}
\STATE \textbf{Input:} Set system $(\mathcal{U},\mathcal{S})$, covering requirement $\rho$, and privacy parameters $(\epsilon,\delta)$
\STATE \textbf{let} $\mathcal{U}_1\leftarrow \mathcal{U}$, $\mathcal{S}_1\leftarrow\mathcal{S}$, $\epsilon^\prime\leftarrow\frac{\epsilon}{2\ln(e/\delta)}$.
\FOR{$i=1,\ldots,m$} 
\STATE \textbf{pick} set $S\in\mathcal{S}_i$ with probability $\propto\exp(\epsilon^\prime |S\cap \mathcal{U}_i|)$.
\STATE \textbf{let} $\pi_i\leftarrow S$,  $\mathcal{U}_{i+1}\leftarrow \mathcal{U}_i-S$, $\mathcal{S}_{i+1}\leftarrow \mathcal{S}_i-\{S\}$.
\ENDFOR
\STATE
\STATE \textbf{let} $T\leftarrow\rho n+ \frac{12\log{m}}{\epsilon}$, $\hat T\leftarrow T+\text{Lap}(2/\epsilon)$.
\FOR{$i=1,\ldots m$} 
\STATE \textbf{let} $f_i\leftarrow |\pi_1\cup\cdots\cup\pi_i|$, $\gamma_i\leftarrow f_i+\text{Lap}(4/\epsilon)$.
\ENDFOR
\STATE \textbf{let} $k$ be first index such that $\gamma_k\ge \hat T$.
\STATE 
\STATE \textbf{Output:} $(\pi,k)$\hfill //$\pi_1,\ldots,\pi_k$ is a set cover for $(\mathcal{U},\mathcal{S})$
\end{algorithmic}
\label{alg:greedy}
\end{algorithm}

\begin{lemma}
Let $k^*$ be the first index such that $\pi_1,\ldots,\pi_{k^*}$ is a $\left(\rho n+\frac{24\log{m}}{\epsilon}\right)$-partial covering of $\mathcal{U}$. Then with probability $1-O(\frac{1}{m})$, for $m=O(n)$, we have $k^*\le O\left(\frac{\ln(m)^2}{\epsilon^\prime(1-\rho)}\right)\cdot\text{OPT}=O\left(\frac{\ln(m)^2\ln(1/\delta)}{\epsilon(1-\rho)}\right)\cdot\text{OPT}$. \label{thm:utility-helper}
\end{lemma}
\begin{proof}
For iteration $i\in[m]$, let $L_i$ be the size of the set which covers the most additional elements (i.e., $L_i=\max_{S\in\mathcal{S}_i}|S\cap \mathcal{U}_i|$). For an iteration where $L_i\ge 6\,\ln{m}/\epsilon^\prime$, the probability of selecting a set which covers less than $L_i-3\,\ln{m}/\epsilon^\prime$ is at most $\frac{1}{m^2}$. Hence, over all iterations where $L_i\ge 6\,\ln{m}/\epsilon^\prime$, we will choose a set which covers at least $L_i/2$ elements with probability at least $1-\frac{1}{m}$. By a standard argument \cite{williamson_shmoys_2011}, this will only use at most $\text{OPT}\cdot \ln{n}$ sets. Consider the last iteration $t$ such that $L_{t}\geq 6\,\ln{m}/\epsilon^\prime$. If the number of elements covered through iteration $t$ is at least $\rho^\prime n$, then we are done. The rest of the proof deals with the case where less than $\rho^\prime n$ are covered.

Next, we analyze what happens when $L_j<6\,\ln{m}/\epsilon^\prime$ for $j=t+1,\ldots,m$. The utility guarantees of the exponential mechanism are essentially useless from this iteration onwards. Notice that the number of remaining elements $|U_j|$ is at most $\text{OPT}\cdot |L_j|$. Unfortunately, we cannot claim that each set chosen covers at least one element; this is simply not true. Instead, we analyze the probability that a set covering at least one element is chosen. Let $\rho^\prime=\frac{\rho+1}{2}$ and note that $\rho^\prime<1$ and $\rho^\prime n\ge \rho n+\frac{24\log{m}}{\epsilon}$ for sufficiently large $n$. Since there are at least $(1-\rho^\prime) n$ uncovered elements remaining and $m=O(n)$, there necessarily exists some constant $\rho^{\prime\prime}$ such that the probability of not covering anything is at most $[1-(1-\rho^{\prime\prime})]$. Thus, the probability of not covering anything over the course of $\frac{2\ln{m}}{1-\rho^{\prime\prime}}$ iterations is at most $$[1-(1-\rho^{\prime\prime})]^{\frac{2\ln{m}}{1-\rho^{\prime\prime}}}\le \exp(-2\ln{m})=\frac{1}{m^2},$$
where we used $1-x\le \exp(-x)$. Thus, each of the $|U_j|$ remaining elements is covered using at most $\frac{2\ln{m}}{1-\rho^{\prime\prime}}$ sets, with probability at least $1-\frac{1}{m}$. Since there are at most $\text{OPT}\cdot |L_j|$ elements remaining which need to be covered, at most $\text{OPT}\cdot \frac{2\ln(m)^2}{\epsilon^\prime (1-\rho^{\prime\prime})}$ sets are used. 
\end{proof}

\begin{lemma}
(Lemma A.1) With probability $1-O\left(\frac{1}{m}\right)$, the threshold $k$ satisfies $|\pi_1\cup\cdots\cup\pi_k|\in\left[\rho n,\rho n+24\log{m}/\epsilon \right]$.
\label{lemma:above}
\end{lemma}

\begin{theorem}
For $\epsilon\in(0,1)$, $\delta<\frac{1}{e}$, and $m=O(n)$, the following are true for Partial Set Cover
\begin{itemize}
    \item Algorithm~\ref{alg:greedy} preserves ($2\epsilon,\delta$)-differential privacy.
    \item With probability $1-O(\frac{1}{m})$, Algorithm~\ref{alg:greedy} is an $O\left(\frac{\ln(m)^2\ln(1/\delta)}{\epsilon(1-\rho)}\right)$-approximation algorithm.
\end{itemize}\label{thm:greedy_main}
\end{theorem}
\begin{proof}
Let's first consider the privacy. Outputting the permutation of sets was shown to be $(\epsilon,\delta)$-differentially private in \cite{gupta2009differentially}. Our mechanism for outputting the threshold $k$ can be viewed as an offline implementation of the AboveThreshold mechanism from \cite{dwork_and_roth}. Since switching to a neighboring set system changes the number of elements covered by a family of sets by at most 1, the analysis of \cite{dwork_and_roth} applies and outputting the threshold is $(\epsilon,0)$-differentially private. By basic composition of adaptive mechanisms \cite{dwork_and_roth}, we have $(2\epsilon,\delta)$-differential privacy.

Now, we turn to the utility guarantee. Consider the threshold $k$ selected; by Lemma~\ref{lemma:above}, the threshold is such that $|\pi_1\cup\cdots\cup\pi_k|$ in the interval $\left[\rho n,\rho n+\frac{24\log{m}}{\epsilon} \right]$ with probability at least $1-\frac{1}{m}$. Hence, it is a $\rho n$-partial cover and by Lemma~\ref{thm:utility-helper}, uses at most $O\left(\frac{\ln(m)^2\ln(1/\delta)}{\epsilon(1-\rho)}\right)\cdot\text{OPT}$ sets.
\end{proof}

\begin{remark}
In Theorem~\ref{thm:greedy_main} (and in future Theorems~\ref{thm:cov} and~\ref{thm:facility})), we can reduce the probability of failure to an arbitrary polynomial in $m$ by losing constant factors in the approximation guarantee. 
\end{remark}

\subsection{Algorithm via Maximum Coverage}

In this section, we give another algorithm for Partial Set Cover under differential privacy. Our algorithm here will give a pseudo-approximation for the problem (i.e., our approximation factor will be with respect to the optimal Set Cover solution instead of the optimal Partial Set Cover solution). Such a guarantee is similar to the one given in \cite{DBLP:conf/icalp/HsuRRU14}. To give our algorithm, we first need to define the Differentially Private Maximum Coverage problem:
\begin{definition}
Let $\mathcal{U}=\{u_1,\ldots,u_n\}$ be the universe of elements and let $\mathcal{S}=\{S_1,\ldots,S_m\}$ be a set system where each $S_i\subseteq \mathcal{U}$ and $\bigcup_{i=1}^{m}S_i=\mathcal{U}$. Finally, let $k$ be the budget. The Maximum Coverage problem asks us to find a size $k$ subset $\{\pi_1,\ldots,\pi_k\}$ of $\mathcal{S}$ such that $|\bigcup_{i=1}^{k}\pi_i|$ is maximized. 
\end{definition}
As in the Partial Set Cover problem, we view the elements of the universe as the private information and we view two set systems as neighbors if they differ by exactly one element $u$ in the universe.
Since the objective here is submodular and monotone~\cite{williamson_shmoys_2011}, we can apply the following result from~\cite{DBLP:conf/icml/MitrovicB0K17} for submodular maximization. We remark that the result we state is stronger than the one given in~\cite{DBLP:conf/icml/MitrovicB0K17}, since we can use a specialized privacy analysis for maximum coverage (like in~\cite{gupta2009differentially}) which is not possible for general submodular functions. This has shown up in other works such as~\cite{DBLP:conf/aaai/JonesNN21}.

\begin{lemma}
There exists an $(\epsilon,\delta)$-differentially private algorithm for the maximum coverage problem which such that the expected number of elements covered is $\left(1-\frac{1}{e}\right)OPT-\frac{2k\ln{n}}{\epsilon_0}$, where $\epsilon_0=\frac{\epsilon}{2\ln(e/\delta)}$.\label{thm:cov}
\end{lemma}

The main idea for our algorithm for Partial Set Cover is that under some restrictions on the set system and budget, the Maximum Coverage problem can be approximated within a constant factor under differential privacy via the algorithm in Lemma~\ref{thm:cov}. Then, iteratively applying the algorithm for the maximum coverage problem with budget $k$ set to the size of the optimal Partial Set Cover suffices to obtain a good approximation algorithm for Partial Set Cover. 

\begin{lemma}
  (Lemma~A.2)
There exists an $(2\epsilon,\delta)$-differentially private algorithm for the maximum coverage problem such that for some constant $C$, if we have
$$k\le \frac{C\epsilon_0}{\ln^2(n)}\cdot\text{OPT},$$
then the algorithm is a $0.15$-approximation with probability $1-O(\frac{1}{n})$, where $\epsilon_0=\frac{\epsilon}{2\ln(e\ln(n)/\delta\ln(1+\alpha))}$. \label{lemma:maxcoverage}
\end{lemma}
Given this result, we can state our algorithm for Partial Set Cover. For simplicity of notation, let's denote the Algorithm referenced in Lemma~\ref{lemma:maxcoverage} by $\textsc{MaxCover}(\mathcal{U},\mathcal{S},k,\epsilon,\delta)$. As mentioned before, the idea is to guess the size of the optimal Partial Set Cover via binary search. Then, assuming we have $\text{OPT}$, we can run $\textsc{MaxCover}$ approximately $O(\log(1-\rho))$ times in order to cover $\rho n$ elements. Though the idea is very simple, there are many subtleties in the algorithm due to privacy. First, our binary search must guarantee that our guess $\text{OPT}^\prime$ is at most $\text{upper}$; this is because the guarantee in Lemma~\ref{lemma:maxcoverage} only applies when the budget isn't too large. Additionally, when binary searching for OPT, we need to decide if the guess is too large or too small, based on the number of elements covered by the output. However, we cannot do this directly since the elements are considered private; as a result, we need to add Laplace noise before making the comparison. This makes the analysis slightly more complicated.
\begin{algorithm}
\caption{$\mathcal{M}_{MaxCoverage}(\mathcal{U},\mathcal{S},\rho,\epsilon,\delta)$}
\begin{algorithmic}
\STATE \textbf{Input:} Set system $(\mathcal{U},\mathcal{S})$, covering requirement $\rho$, privacy parameters $(\epsilon,\delta)$
\STATE \textbf{let} $\text{upper}=\lfloor\frac{C(1-\frac{\rho}{2})n\epsilon_0}{\ln^3(n)}\rfloor$, $t=\lceil\log_{0.85}(1-\rho^\prime)\rceil$.
\STATE \textbf{Binary Search} on $\{1,\ldots,\text{upper}\}$, and let the current guess be $\text{OPT}^\prime$
\STATE \quad\textbf{let} $\text{SOL}=\emptyset$, $\mathcal{U}_1\leftarrow \mathcal{U}$, $\mathcal{S}_1\leftarrow\mathcal{S}$\STATE \quad\textbf{let} $\epsilon^\prime\leftarrow\frac{\epsilon}{t\log_2(n)},\delta^\prime\leftarrow\frac{\delta}{t\log_2(n)}$.  
\STATE \quad\textbf{for} {$i=1,\ldots,t$} \textbf{do}
\STATE \quad\quad\textbf{run} $\textsc{MaxCover}(\mathcal{U}_i,\mathcal{S}_i,\text{OPT}^\prime,\epsilon^\prime,\delta^\prime)$ to obtain sets $\pi_i=\{\pi_{i,1},\ldots,\pi_{i,\text{OPT}^\prime}\}$.
\STATE \quad\quad\textbf{let} $\text{SOL}\leftarrow\text{SOL}\cup \pi_i$
\STATE\quad\quad \textbf{let} $\mathcal{U}_{i+1}\leftarrow \mathcal{U}_i-\bigcup_{j=1}^{\text{OPT}^\prime}\pi_{i,j}$, $\mathcal{S}_{i+1}\leftarrow \mathcal{S}_i-\pi_i$.
\STATE \quad\textbf{endfor}
\STATE \quad\textbf{let} $\gamma$ be the number of elements covered by SOL
\STATE \quad\textbf{let} $\hat\gamma=\gamma+\frac{\log{n}}{\epsilon^\prime}+\text{Lap}(1/\epsilon^\prime)$
\STATE \quad \textbf{if} $\hat\gamma\ge\rho n$ \textbf{increase} $\text{OPT}^\prime$; otherwise, \textbf{decrease} $\text{OPT}^\prime$

\STATE \textbf{Output:} $\text{SOL}$ for minimum $\text{OPT}^\prime$ satisfying $\hat\gamma\ge\rho n$
\end{algorithmic}
\label{alg:iterate}
\end{algorithm}

\begin{theorem}\label{thm:alg2}
Algorithm~\ref{alg:iterate} is $(\epsilon,\delta)$-DP. Furthermore, assuming the optimal set cover has size $\text{OPT}\le\frac{C(1-\frac{\rho}{2})n\epsilon_0}{\ln^3(n)}$, where $C$ is from Lemma~\ref{lemma:maxcoverage} and $\epsilon_0=O\left(\frac{\epsilon/t}{\ln\ln(n)+\ln(t/\delta)}\right)$, Algorithm~\ref{alg:iterate} outputs a $(1-\rho)$-Partial Set Cover using at most $O(\log(\frac{1}{1-\rho}))\cdot\text{OPT}$ sets with probability $1-\tilde O(\frac{1}{n})$.
\end{theorem}
\begin{proof}
First, let's consider the privacy guarantee. For each iteration of the binary search, we run \textsc{MaxCover} $t$ times. By basic composition, we this is $(t\epsilon^\prime,t\delta^\prime)$-differentially private. Additionally, $\hat\gamma$ is the output of the Laplace Mechanism, so it is $(t\epsilon^\prime,0)$-differentially private. The binary search takes at most $\log_2(\text{upper})\le\log_2(n)$ iterations to converge, so $(2\epsilon,\delta)$-differential privacy follows by basic composition once again. Finally, the outputted solution $\text{SOL}$ is $(2\epsilon,\delta)$-differentially private by post-processing.

Now, we will turn to the utility guarantee. Let $\rho^\prime=\frac{\rho+1}{2}$, $\beta=0.15$, and let us first consider the algorithm when our guess $\text{OPT}^\prime$ is at least $\text{OPT}$. We claim that running \textsc{MaxCover} $t$ times, as in Algorithm~\ref{alg:iterate}, covers at least $\rho^\prime n\ge \rho n + \frac{2\log{n}}{\epsilon^\prime}$ elements. Since the output of the Laplace mechanism $\hat\gamma$ will be at least $\rho n$ with probability at least $1-\frac{1}{n}$, the binary search will converge to some $\text{OPT}^\prime\le\text{OPT}$. Now, consider the partial set cover output by the algorithm; we know that the number of sets used is $t\cdot \text{OPT}^\prime\le t\cdot\text{OPT}$, hence a $t$-approximation. Next, we will prove our claim.

If at any iteration $i<t$, we have $\rho^\prime n$ elements are covered by the previously selected sets, we are done. Suppose there remains at least $(1-\rho^\prime)n$ elements uncovered at all iterations $i\le t$; we will show that $\rho^\prime n$ elements are covered after iteration $t$. By definition of $\text{OPT}$ and the fact that $\text{OPT}^\prime\ge\text{OPT}$, there exists $\text{OPT}^\prime$ sets which cover the remaining uncovered elements. Thus, when $\text{OPT}$ is suitably small as in the theorem statement,  running \textsc{MaxCover} always covers at least an $\alpha$-fraction of the remaining elements, leaving a $(1-\alpha)$-fraction of the remaining elements uncovered. By algebra, running \textsc{MaxCover} t times suffices to guarantee that at most $(1-\rho^\prime)n$ elements remain uncovered.
\end{proof}

\begin{remark}
We emphasize that the guarantees in Theorem \ref{thm:alg2} are pseudo-approximations. That is, the approximation guarantee is with respect to the optimal set cover whereas the algorithm outputs a \emph{partial} set cover.
\end{remark}
\section{An Application to Facility Location}\label{sec:fac}

To show an example where an \emph{explicit} solution for (partial) set cover is a necessary building block for a larger algorithm, we will consider a facility location problem called \prob{}, introduced in \cite{li2022deploying}. The authors introduced the following generalization of the well known $k$-supplier problem for deploying vaccine distribution sites:
\begin{definition}
Let $\mathcal{C}$ be a set of locations in a metric space with distance function $d: \mathcal{C} \times \mathcal{C} \mapsto \mathbb{R}_{\geq 0}$. Let $\mathcal{P}$ be a set of $n$ people where each person $p \in \mathcal{P}$ is associated with a set $S_p \subseteq \mathcal{C}$, which can be interpreted as the set of locations $p$ visits throughout the day. Finally, let $k\in\mathbb{N}$ be a budget on the number of facilities. We want to output a set of locations $F \subseteq \mathcal{C}$ with $|F|\le k$ to place facilities which minimizes $\max_{p\in\mathcal{P}}d(S_p, F)$, where $d(S, F) = \min_{j \in S, j' \in F}d(j,j')$. For simplicity, we normalize the metric so that the diameter is $\max_{j,j^\prime\in \mathcal{C}}d(j,j^\prime)=1$.
\label{def:vacc}
\end{definition}

We consider the outliers version of the above problem, where we are only required to serve $\lfloor\rho n\rfloor$ people in the population, for some $\rho<1$. Since the problem was designed for vaccine distribution, the outliers variant is still very interesting: we only need to vaccinate $94\%$ of the population in order to achieve herd immunity. For the differential privacy, the set of locations $\mathcal{C}$, covering requirement $\rho$, and budget $k$ is public information and the individuals along with their travel patterns is private information. We call two instances of \prob{} neighbors if they differ by exactly one individual $p$ (along with their travel pattern $S_p$). Thus, we protect the privacy of the individuals: a differentially private algorithm's output will not differ \emph{too much} whether or not any individual and their travel patterns are included in the data.

Unfortunately, we show that this problem is hard in a very strong sense: there cannot exist any (efficient or inefficient) algorithm which provides a finite approximation guarantee. Hence, we turn towards bicriteria algorithms where we violate the budget $k$: an algorithm is an $(\alpha,\beta)$-bicriteria approximation for \prob{} with outliers if it obtains an additive $\beta$-approximation to the optimal radius while placing at most $\alpha k$ facilities. We show that this is possible. The idea of the algorithm is simple: we first guess the optimal radius $R^*$ via a binary search on the interpoint distances (in practice, we just guess $R^*$ up to an additive error of $\gamma$ for faster convergence). Assuming we know $R^*$, we consider the \emph{reverse} problem where we wish to place the fewest facilities in order to cover a $\rho$ fraction of the clients within a radius of $R^*$. This is exactly a Partial Set Cover problem, so we can apply Algorithm~\ref{alg:greedy} (it is easy to verify the privacy requirements of \prob{} matches that of Partial Set Cover). To formalize this, let the universe be $\mathcal{U}=\mathcal{P}$ and let the set system be $\mathcal{S}_R=\{S_{j}(R):j\in\mathcal{C}\}$, where $S_{j}(R)=\{p\in\mathcal{P}:d(S_p,j)\le R\}$. It is easy to see that this indeed reduces to a Partial Set Cover problem. In the following formalization of our algorithm, let $\alpha(\epsilon,\delta)$ denote the approximation guarantee of Algorithm \ref{alg:greedy}:

\begin{algorithm}
\caption{\textsc{DPClientCover:}}
\begin{algorithmic}
\STATE \textbf{Input:} \prob{} instance $(\mathcal{C}, \mathcal{S}, \mathcal{P}, k, \rho)$, additive error $\gamma$, and privacy parameters $(\epsilon,\delta)$
\STATE \textbf{let} $\epsilon^\prime\leftarrow{\epsilon}/{\log_{2}(1/\gamma)}, \delta^\prime\leftarrow{\delta}/{\log_{2}(1/\gamma)}$
\STATE \textbf{let} $low \leftarrow 0$, $high \leftarrow 1$
\WHILE {$high - low > \gamma$}
\STATE $\R \leftarrow (high+low)/2$
\STATE Calculate $\mathcal{S}_\R$ as described above
\STATE $F_R\leftarrow \mathcal{M}_{PartialSetCover}(\mathcal{C},\mathcal{S}_R,\rho,\epsilon',\delta')$
\STATE\textbf{if} {$|F_R| > \alpha(\epsilon^\prime,\delta^\prime) \cdot k$} \textbf{then} $low \leftarrow \R$ \\
\STATE\textbf{otherwise} $high \leftarrow \R$
\ENDWHILE
\STATE \textbf{Output} $F_R$ for minimum $R$ such that $|F_R| \leq \alpha(\epsilon^\prime,\delta^\prime) \cdot k$.
\end{algorithmic} \label{alg:client-cover}
\end{algorithm}

\begin{theorem}
For any $\gamma>0$, Algorithm~\ref{alg:client-cover} is $(2\epsilon,\delta)$-DP and an $(O\left(\frac{\log^2|\mathcal{C}|\log(1/\gamma)\log(1/\delta)}{\epsilon}\right),\gamma)$-approximation algorithm for \prob{} with outliers, with probability at least $1-\tilde O\left(\frac{1}{|\mathcal{C}|}\right)$, when $|\mathcal{C}|=O(n)$.
\label{thm:facility}
\end{theorem}
\begin{proof}
The privacy guarantee is easy: since the guesses $R$ don't depend on the private information, $(2\epsilon,\delta)$-differential privacy follows directly by basic composition. Note that we are allowed to find the minimum $R$ such that $|F_R|\le \alpha\cdot k$ by post-processing, since the $F_R$'s are private already.

Next, we turn to the utility guarantee. Let us analyze the algorithm for the iterations of binary search where $R\ge R^*$. We will show that for these iterations, we necessarily have $|F_R|\le\alpha(\epsilon^\prime,\delta^\prime)\cdot k$. This implies that the binary search converges to some $R$ satisfying $R\le R^*+\gamma$, so the minimum $R$ satisfying $|F_R|\le \alpha(\epsilon^\prime,\delta^\prime)\cdot k$ also will satisfy $R\le R^*+\gamma$.

To complete the proof, let's prove our claim. Suppose $R\ge R^*$ and consider the corresponding Partial Set Cover instance. We know there exists some Partial Set Cover of size $k$ (by the definition of $R^*$), so our approximation guarantees in Theorem \ref{thm:greedy_main} imply that $|F_R|\le\alpha(\epsilon^\prime,\delta^\prime)\cdot k$ with probability $1-O(\frac{1}{n})$. Apply the union bound over all iterations of the binary search completes the proof.
\end{proof}

\begin{remark}
Our algorithm needs to violate the budget $k$ by a poly-logarithmic multiplicative factor, which is often not possible in the real world. To circumvent this, we note that it has been observed experimentally that set cover algorithms obtain near optimal solutions; thus, we can implement Algorithm~\ref{alg:client-cover} with $\alpha$ set to $1$ and still obtain a near-optimal radius $R$ in practice. Even this contribution is non-trivial; before our work, even good heuristics were not known for differentially private $k$-supplier. We use this heuristic in the experiments in Section \ref{sec:exp}.
\end{remark}

\begin{remark}
We state Algorithm \ref{alg:client-cover} and its guarantees in terms of guessing the optimal radius $R^*$ up to an additive error of $\gamma$. We find that this converges faster in practice and thus obtains a better privacy/utility tradeoff. As mentioned before, binary searching on the interpoint distances can save us this additive error of $\gamma$ but will lose a logarithmic factor in the number of facilities placed. The proof of this is very similar, and thus omitted.
\end{remark}

\subsection{Lower Bounds}

To give our lower bound for \prob{}, we will first show that Partial Set Cover cannot be solved exactly under $(\epsilon,\delta)$-differential privacy and that an approximation is necessary (deferred to Appendix).
\begin{lemma}
Under $(\epsilon,\delta)$-differential privacy, Partial Set Cover cannot be solved exactly (with high probability).\label{thm:lower-bound}
\end{lemma}

We will use this fact to derive an information theoretic lower bound for \prob{}, stating that even computationally inefficient algorithms cannot give a finite approximation factor for this problem while simultaneously satisfying approximate differential privacy. Overall, this lower bound justifies our use of bicriteria approximation algorithms for our problem. The reduction we give is similar to the one found in \cite{DBLP:journals/mp/AneggAKZ22} for a separate problem, $\gamma$-colorful $k$-center.

\begin{theorem}
There can be no finite approximation algorithm for \prob{} which satisfies $(\epsilon,\delta)$-differential privacy, even when the metric space is the Euclidean line.
\end{theorem}
\begin{proof}
Suppose for contradiction some differentially private approximation algorithm did exist with approximation ratio $\alpha=\alpha(n,m,\rho,\epsilon,\delta)$. We will show that it can be used to solve any instance $(\mathcal{U},\mathcal{S},\rho,\epsilon,\delta)$ of Differentially Private Partial Set Cover to optimality. By contradiction, we will conclude such an approximation algorithm cannot exist.

Let the metric space be the Euclidean line, let $\mathcal{C}=\{1,\ldots,|\mathcal{S}|\}$ and let the covering requirement be $\rho=\rho$. For each element $u_\ell\in\mathcal{U}$, define a person $p\in\mathcal{P}$ and let their travel locations be $S_p=\{i\in[|\mathcal{S}|]:u_\ell\in S_i\}$. By construction, it is clear that the optimal solution for this \prob{} instance has radius $0$ and gives a solution (of size $k$) to the Partial Set Cover instance. Furthermore, any $\alpha$-approximation will output a solution with radius $\alpha\cdot 0=0$ as well. Consequently, we can binary search over $k\in[m]$ to use any $\alpha$-approximation for \prob{} to obtain an optimal solution for Partial Set Cover.
\end{proof}

\section{Experiments}\label{sec:exp}
\begin{table}[h]
    \centering
    \begin{tabular}{c|cccc}
    \toprule
     & Clients  & Locations  & Diameter (km) \\
    \midrule
    Charlottesville & 33156 & 5660 & 8.12\\
    Albemarle & 74253 & 9619 & 61.62\\
    \bottomrule
    \end{tabular}
    \caption{Network Information}
    \label{tab:my_label}
\end{table}
For our experiments in our main paper, we focus on the vaccine distribution problem we studied in Section~\ref{sec:fac}. We run all of our experiments on synthetic mobility data from Charlottesville city and Albemarle county, which are part of a synthetic U.S. population (see~\cite{chen2021,machi2021} for details). The data keeps track of the locations each person travels to throughout a week, so each person is associated with a set of activity locations. We use this data in our experiments to study the following:
\begin{itemize}
    \item \textbf{Privacy/Accuracy Tradeoff:} how does the quality of our algorithm vary as the privacy parameters $(\epsilon,\delta)$ vary?
    \item \textbf{Cost of Privacy:} how does our private algorithm compare with the non-private one given in \cite{li2022deploying}?
\end{itemize}
Due to space constraints, we defer some of these experiments and analyses to the full version. There, we also investigate the above questions for our Partial Set Cover algorithms.

\textbf{Baseline:} We use a non-private greedy algorithm to solve the partial set cover problem as our \emph{Baseline}. Starting with an empty set of locations, the greedy algorithm at each step chooses one location that covers the most of the uncovered clients. Then it adds the location to its set and updates the current state of covered clients. The baseline algorithm utilizes the same binary search routines with~\textsc{DPClientCover} to find the optimal radius. 

\textbf{Metric:} All of our comparisons are made based on the objective value of \prob{} with outliers. Informally, by calculating the service cost of each person, our objective is to minimize the distance the $\lceil\rho n\rceil^{th}$ person needs to travel in order to get vaccinated. 
Formally, our objective is $\min_{P_1\subseteq \mathcal{P}:|P_1|=\lceil\rho n\rceil}\max_{p\in P_1}d(S_p,F)$.

\subsection{Privacy/Accuracy/Budget Tradeoff}
In this subsection, we investigate the two questions regarding the privacy/accuracy trade-off and the cost of privacy. We set $\delta=10^{-6}$ and run our private vaccine distribution algorithm for different budgets $k\in\{4,\ldots,16\}$ and different privacy parameters $\epsilon\in\{0.25,0.5,1,2,4\}$.  A careful reader may notice that Theorem \ref{thm:greedy_main} only applies for $\epsilon\in(0,1)$, and may worry that our experimental regime doesn't have privacy guarantees. However, \textsc{DPClientCover} makes calls to the Partial Set Cover algorithm with parameter $\epsilon^\prime=\epsilon/\log_2(1/\gamma)$, so we still have differential privacy.

\begin{figure}[h!]
    \centering
    \includegraphics[width=\columnwidth,keepaspectratio]{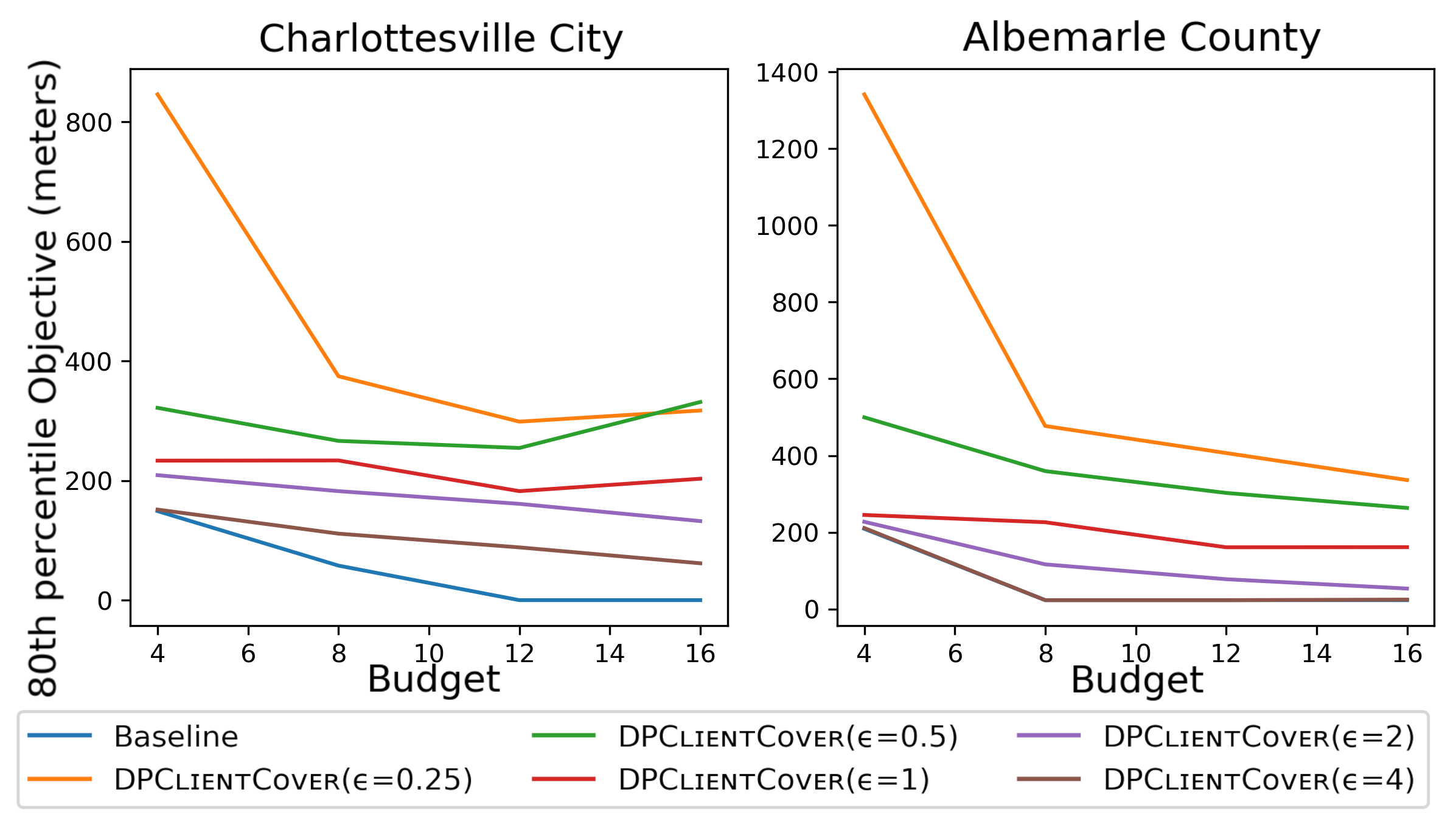}
    \caption{Utilities of \textsc{DPClientCover} at $\rho=0.8$ with different values of privacy parameter ($\epsilon$) and budget ($k$) on both datasets. The $y$-axis indicates the objectives measured at the $80^{th}$ percentiles (in meters). The $x$-axis shows different budget values, i.e., the number of selected locations. Generally, \textsc{DPClientCover} performs closely to the \emph{Baseline} with large values of $\epsilon$ ($\epsilon \geq 2$) on both datasets. For the Albemarle county dataset, curves for the \emph{Baseline} and \textsc{DPClientCover} with $\epsilon =4$ overlap.}
    \label{fig:budget-tradeoff}
  \end{figure}

The experimental results show that, generally, with higher values of budget, all algorithms perform better, i.e., they achieve lower objective values. Our private algorithm \textsc{DPClientCover} usually achieves higher accuracy with higher $\epsilon$ at every level of budget, while the non-private algorithm (\emph{Baseline}) always has the best performance. These results are expected. Figure~\ref{fig:budget-tradeoff} also shows that our private algorithm performs well, obtaining near optimal choices of facility locations for $\epsilon\ge 1$. For example, \textsc{DPClientCover} with $\epsilon=4$ matches the performance of \emph{Baseline} on the Albemarle dataset and almost matches the Baseline at several values of budget on the Charlottesville dataset.
However, we can observe some trade-offs between privacy and utility, especially in high-privacy settings. In particular, for $\epsilon=0.25$ and $k=4$, there is a high cost of guaranteeing privacy in both datasets. In these cases, the objectives of the private settings are 4--7 times higher than those of the Baseline.

A more subtle result that Figure~\ref{fig:budget-tradeoff} gives is to help us understand the trade-off between the budget and the privacy loss $\epsilon$.
From a policy-making perspective, this is interesting since we can allocate more mobile vaccine distribution sites in return for a stronger privacy guarantee. For example, the accuracy from $k=8$ facilities with $\epsilon=0.25$ is similar to using $k=4$ facilities with $\epsilon=0.5$. Figure~\ref{fig:Cville} shows the vaccine clinics chosen by the private ($\epsilon=0.5$, white spots) and non-private (black spots) algorithms for the Charlottesville dataset with $k=8$. The two algorithms tend to choose two totally different distributions of vaccine clinics, although they do not differ much in term of objective values (see Figure~\ref{fig:budget-tradeoff}, where $k=8$).

\section{Discussion}
In this paper, we consider the Set Cover problem and the $k$-supplier problem, which have strong hardness results under differential privacy. We observe that partial coverage of elements/clients suffices to avoid the impossibility results, and give the first non-trivial approximation algorithms for both problems. Overall, our work is an important step in tackling the impossibility results in differentially private algorithm design and leaves many interesting problems open:
\begin{itemize}
    \item Both of our algorithms for Partial Set Cover require some (relatively loose) assumption on the set system. An interesting question is whether we can remove these assumptions: can we obtain a general approximation algorithm for Partial Set Cover under differential privacy?
    \item Our algorithm for $k$-supplier violates the budget $k$ by a poly-logarithmic factor, which is impractical in some settings. It is interesting to see what guarantees are possible without violating the budget: can we obtain true approximations for $k$-supplier with outliers with privacy?
    \item As mentioned before, the facility location problem has a $\Omega(\sqrt{n})$ approximation hardness result under differential privacy. It is interesting to see if our ideas can help avoid this: does allowing partial coverage circumvent hardness results of the uniform facility location problem?
\end{itemize}

\section*{Acknowledgements}
Dung Nguyen and Anil Vullikanti were partially supported by 
NSF grants CCF 1918656, IIS 1955797,  NIH grant 2R01GM109718-07, and
CDC cooperative agreement MIND U01CK000589. George Li was supported in part by NSF award number CCF-1918749.

\bibliographystyle{named.bst}
\bibliography{refs.bib}

\appendix

\onecolumn

\section{Missing Proofs}

\begin{lemma} (Lemma~2.3)
With probability $1-O\left(\frac{1}{m}\right)$, the threshold $k$ is such that $|\pi_1\cup\cdots\cup\pi_k|\in\left[\rho n,\rho n+\frac{24\log{m}}{\epsilon} \right]$.
\label{lemma:above-appendix}
\end{lemma}

\begin{proof}
Let $A$ be the last index such that $f_A\le \rho n$ and let $B$ be the first index such that $f_B\ge \rho n+\frac{24\log{m}}{\epsilon}$. Suppose $k$ doesn't satisfy the requirements in the theorem statement; then either (i) for some $1\le i\le A$, we have $\gamma_i\ge \hat T$ or (ii) for some $B\le j\le m$, we have $\gamma_j\le \hat T$. 
We will bound the probability that these events occur. Let $1\le i\le A$; we have the following
\begin{align}
    \Pr[\gamma_i\ge \hat T]&=\Pr\left[\text{Lap}(2/\epsilon)+\text{Lap}(4/\epsilon)\ge \frac{12\log{m}}{\epsilon}\right]\nonumber\\
    &\le \Pr\left[\text{Lap}(2/\epsilon)\ge\frac{4\log{m}}{\epsilon}\right]+\Pr\left[\text{Lap}(4/\epsilon)\ge \frac{8\log{m}}{\epsilon}\right]\le \frac{2}{m^2}\nonumber
\end{align}
Then by the union bound, the probability that (i) occurs is at most $O(\frac{1}{m})$. The bound for (ii) is similar (and in fact, symmetric) so the lemma follows directly.
\end{proof}

\begin{lemma}
  (Lemma~2.7)
There exists an $(2\epsilon,\delta)$-differentially private algorithm for the maximum coverage problem such that for some constant $C$, if we have
$$k\le \frac{C\epsilon_0}{\ln^2(n)}\cdot\text{OPT},$$
then the algorithm is a $0.15$-approximation with probability $1-O(\frac{1}{n})$, where $\epsilon_0=\frac{\epsilon}{2\ln(e\ln(n)/\delta\ln(1+\alpha))}$. \label{lemma:maxcoverage-appendix}
\end{lemma}

\begin{proof}
Let $\alpha<1-\frac{1}{e}$ be a small constant and take $C=(1-\frac{1}{e}-\alpha)\ln(1+\alpha)/2$. By algebra, we see that if we $k$ is not too large as in the lemma statement, then Theorem~2.6 implies that there exists an $(\epsilon^\prime,\delta^\prime)$-differentially private algorithm for the maximum coverage problem which is an $\alpha$-approximation to the optimal solution in expectation, where $\epsilon^\prime=\frac{\ln(1+\alpha)}{\ln{n}}\epsilon$ and $\delta^\prime=\frac{\ln(1+\alpha)}{\ln{n}}\delta$. Note that the current approximation guarantee for the algorithm is in expectation, but we will need something slightly stronger.

To convert the approximation guarantee from to a guarantee in expectation to one with high probability, we can simply repeat the algorithm $T=\frac{\ln{n}}{\ln(1+\alpha)}$ times and choose the solution which cover the most elements (via the exponential mechanism). Note that repeating the algorithm $T$ times is $(\epsilon,\delta)$-differentially private by basic composition; since the exponential mechanism is $\epsilon$-differentially private, our entire mechanism is $(2\epsilon,\delta)$-differentially private, as desired.

Next, we analyze the utility of our proposed mechanism. Let $X_1,\ldots,X_T$ be the (random) number of elements covered by the sets chosen by the algorithm in $T$ independent runs. Let $i\in[T]$ be arbitrary; by Markov's inequality, we have
\begin{align}
    \Pr[\text{OPT}-X_i\ge (1+\alpha)\expected[\text{OPT}-X_i]]\le\frac{1}{1+\alpha}.\nonumber
\end{align}
Since $\expected[X_i]\ge\alpha\cdot \text{OPT}$, we have $\expected[\text{OPT}-X_i]\le (1-\alpha)\text{OPT}$ so we can claim
\begin{align}
    \Pr[\text{OPT}-X_i\ge (1+\alpha)(1-\alpha)\text{OPT}]=\Pr[\text{OPT}-X_i\ge(1-\alpha^2)\text{OPT}]\le\frac{1}{1+\alpha}.\nonumber
\end{align}
Moving terms around, we can rewrite the above as
\begin{align}
    \Pr[X_i\le \alpha^2\cdot\text{OPT}]\le\frac{1}{1+\alpha}.\label{eq:markov}
\end{align}
Using this, we can conclude
\begin{align}
    \Pr[\max_{i\in[T]}X_i>\alpha^2 \cdot\text{OPT}]=1-\Pr[\max_{i\in[T]}X_i\le \alpha^2\cdot\text{OPT}]=1-\prod_{i=1}^{T}\Pr[X_i\le\alpha^2\cdot\text{OPT}]\ge 1-\frac{1}{n},\nonumber
\end{align}
where the final inequality follows by (\ref{eq:markov}). Finally, we need to apply the exponential mechanism on these $T$ families of sets to guarantee privacy. Let $X$ be the number of elements covered by the chosen set; by the utility guarantees of the exponential mechanism, we have
\begin{align}
    \Pr\left[X\le \max_{i\in[T]}X_i - \frac{4\ln{n}}{\epsilon}\right]\le\frac{1}{n}
\end{align}
Note that by our assumption on $k$, we have $\text{OPT}\ge \frac{k\ln^2(n)}{C\epsilon_0}$. For even moderately large $n$, this implies $0.1\cdot\text{OPT}\ge \frac{4\ln{n}}{\epsilon}$, so we have
\begin{align}
    \Pr\left[X\ge (\alpha^2-0.1)\text{OPT}\right]\ge 1-O\left(\frac{1}{n}\right).
\end{align}
Taking $\alpha=0.5$ suffices to give us a $0.15$-approximation algorithm with high probability. 
\end{proof}

\begin{lemma} (Lemma~3.3) Under $(\epsilon,\delta)$-differential privacy, Partial Set Cover cannot be solved exactly (with high probability). \label{thm:lower-bound-Appendix}
\end{lemma}
\begin{proof}
Let's consider a Partial Vertex Cover instance.
Let $S_n$ denote the star graph on $n$ vertices (i.e., there are $n-1$ nodes all connected to a single center node). Our graph $G$ will consist of two star graphs $S_{n/2}$ (centered at nodes $u$ and $v$) connected together with a single edge $(u,v)$. Let the covering requirement be $\rho=\frac{1}{2}$. Clearly, the optimal Partial Vertex Cover has size 1 by choosing $u$ or $v$; assume for contradiction either $u$ or $v$ is output with probability $1-o(1)$. Now consider any graph $G^\prime$ with $3$ additional edges in $G$. Now the covering constraint $\rho$ requires our solution to cover at least $\frac{n}{2}+2$ edges, so only choosing $u$ or $v$ is insufficient. But by group privacy, the algorithm will still choose either $u$ or $v$ with probability at least $e^{-3}(1+o(1))$, a contradiction.
\end{proof}

\section{Discussion of Privacy Definition}

Recall that when defining differentially private set cover, we consider two Partial Set Cover instances neighboring if the set systems $(\mathcal{U}_1,\mathcal{S}_1)$ and $(\mathcal{U}_2,\mathcal{S}_2)$ are two neighboring datasets if they differ by exactly one element $u$ in the universe and the sets $S_{i1}\in\mathcal{S}_1$ and $S_{i2}\in\mathcal{S}_2$ differ only by $u$ (or are the same). In contrast, the definition in \cite{gupta2009differentially} has the set system $(\mathcal{U},\mathcal{S})$ as public information and a set $\mathcal{R}\subseteq \mathcal{U}$ as the private set of elements which need to be covered. Two set cover instances are neighboring if the underlying set system is the same and $\mathcal{R}_1,\mathcal{R}_2\subseteq\mathcal{U}$ differ by exactly one element. The two definitions are similar in that they both give privacy with respect to the elements in the universe. We claim that in general, our guarantee of privacy is stronger: our privacy implies that of \cite{gupta2009differentially} but the converse is not true. Indeed, observe that any neighboring instance in the definition of \cite{gupta2009differentially} is also a neighboring instance in our definition. Hence, the guarantee of differential privacy applies directly. Conversely, suppose $(\mathcal{U},\mathcal{S})$ is a set system and let $\mathcal{R}_1\subseteq\mathcal{U}$ be the private elements which need to be covered. This can be viewed as Partial Set Cover instance $(\mathcal{U}_1,\mathcal{S}_1)$ where $\mathcal{U}_1=\mathcal{R}_1$ and $\mathcal{S}_1=\{S\cap \mathcal{R}_1: S\in\mathcal{S}\}$. Now suppose a mechanism is differentially private with respect to the definition in \cite{gupta2009differentially}. Then for any $\mathcal{R}_2\subseteq\mathcal{U}$ which differs from $\mathcal{R}_1$ by one element, we have differential privacy. But this does not imply privacy with our definition since our neighboring Partial Set Cover instance $\mathcal{U}_2$ can differ from $\mathcal{U}_1$ by an arbitrary element, not necessarily in $\mathcal{U}$. Hence, our guarantee is in general stronger.

\section{Additional Experimental Results}

    \begin{figure}[h!]
    \centering
    \includegraphics[width=0.5\columnwidth,keepaspectratio]{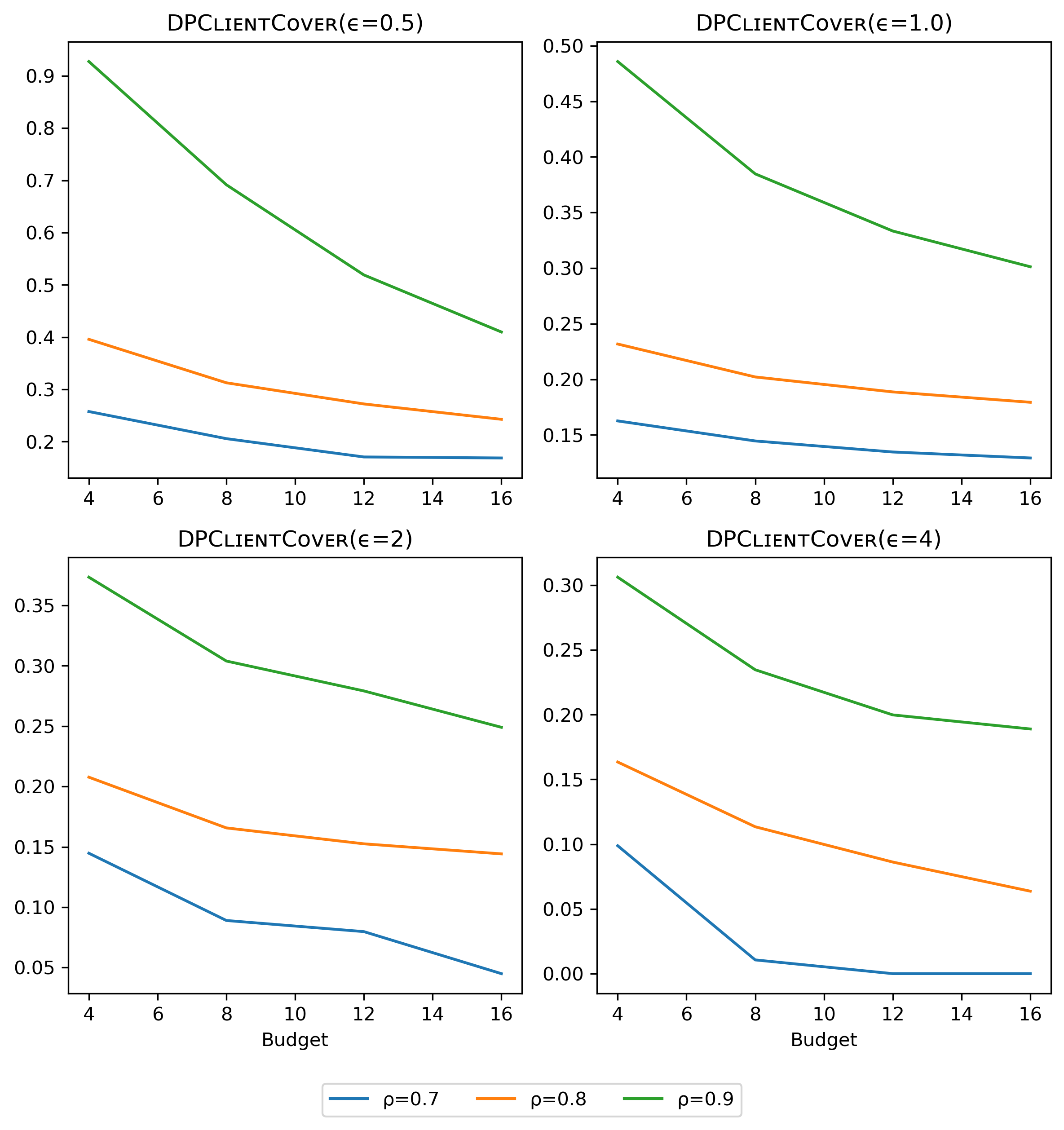}
    \caption{Utilities of \textsc{DPClientCover} with different values of the fraction to cover $\rho$ and privacy budget $\epsilon$ on the Charlottesville city dataset. Utilities are measured as the objectives of the $100*\rho{th}$ percentile of the population for each value of $\rho$. Statistics are reported as the average of $10$ repeats for each combination of parameters.}
    \label{fig:rho-comparison}
  \end{figure}

\end{document}